# Atomic scale evolution of the surface chemistry in Li[Ni,Mn,Co]O$_2$ cathode for Li-ion batteries stored in air

Mahander P. Singh[a], Se-Ho Kim[a,*], Xuyang Zhou[a], Hiram Kwak[b], Stoichko Antonov[c], Leonardo Shoji Aota[a], Chanwon Jung[a], Yoon Seok Jung[b], Baptiste Gault[a,d,*]

a Max-Planck-Institut für Eisenforschung, Düsseldorf, Germany

b Department of Chemical and Biomolecular Engineering, Yonsei University, Seoul, South Korea

c National Energy Technology Laboratory, Albany, Oregon, USA

d Department of Materials, Royal School of Mines, Imperial College London, London, UK

* co-corresponding authors: s.kim@mpie.de; b.gault@mpie.de

**Abstract**

Layered LiMO$_2$ (M = Ni, Co, Mn, and Al mixture) cathode materials used for Li-ion batteries are reputed to be highly reactive through their surface, where the chemistry changes rapidly when exposed to ambient air. However, conventional electron/spectroscopy-based techniques or thermogravimetric analysis fails to capture the underlying atom-scale chemistry of vulnerable Li species. To study the evolution of the surface composition at the atomic scale, here we use atom probe tomography and probed the surface species formed during exposure of a LiNi$_{0.8}$Mn$_{0.1}$Co$_{0.1}$O$_2$ (NMC811) cathode material to air. The compositional analysis evidences the formation of Li$_2$CO$_3$. Site specific examination from a cracked region of an NMC811 particle also suggests the predominant presence of Li$_2$CO$_3$. These insights will help to design improved protocols for cathode synthesis and cell assembly, as well as critical knowledge for cathode degradation

During storage in air under ambient conditions of temperature and pressure, the high surface reactivity of Ni-rich layered oxide cathode materials used in rechargeable Li-ion battery leads to the generation of certain surface species, which causes problems during electrode slurry preparation, cell storage and charge/discharge cycling[1–3]. For example, Jung et al.[4] showed that ambient storage of Ni-rich NMC ($LiNi_xMn_yCo_{1-x-y}O_2$) cathode for a year led to significant surface contamination. The authors suggested a possible formation of Ni carbonate species, making the particles electrochemically inactive with severely degraded electronic conductivity and hindered Li transport. Eventually there was considerable impedance growth and irreversible capacity loss. In another report, Busa et al.[5] reported that NMC particles exposed to humid air for 28 days showed an increase of capacity over the first 25 cycles, suspected to be due to the decomposition of possible lithium carbonates formed during air-exposure, followed by a dramatic capacity fade of 70%, whereas the pristine sample exhibited a stable cycling performance (98% capacity retention). Archival literature suggests possible formation of lithium/nickel hydroxides, oxides and carbonates during air/moisture interaction with NMC surfaces at near room temperatures[4,6]. The chemical composition of the surface species is very important in terms of electrical resistance and Li-ion transport. However, the actual atomic-scale chemistry of Ni-rich oxide cathode materials after exposure to air or moisture remains elusive (nickel *vs* lithium carbonate/hydroxides/oxides). Yet it impacts battery operation, lifetime and the formation and chemistry of the cathode-electrolyte interphase which are subjects of intensive research efforts[7–9]. Elucidating the surface reactions will ultimately allow for improved understanding of the degradation pathways and underpin its solutions.

The chemical and structural evolution of the cathode-air interface is challenging to image and is mostly studied by electron/spectroscopy-based techniques or thermogravimetric analysis.

Spatially-resolved techniques such as scanning/transmission electron microscopy (SEM/TEM) are widely applied to address the surface degradation phenomenon. However, Li-based compounds are susceptible to electron beam damage, which complicates data interpretation[10–13]. Similarly, X-ray illumination of Li containing compounds forces an uncontrollable evolution towards stable lithium oxide, even in the soft X-ray regime, precluding safe and reliable application of X-ray based spectroscopy techniques[14–17]. Alternative analytical techniques are now being explored to circumvent these issues. Time-of-flight secondary ion mass spectrometer (tof-SIMS) and atom probe tomography (APT) which provides quantitative, analytical three-dimensional imaging with sub-nanometer resolution are currently the most suitable solutions[18,19].

Here, we use APT, complemented by cryogenic-electron- and ion-microscopy techniques[20], to track the morphological and compositional evolution of the surface of $Li(Ni_{0.8}Mn_{0.1}Co_{0.1})O_2$ (NMC811) particles, commercial cathode materials, over increasing storage time under ambient air conditions. These insights aid to further understand the relationship between surface chemistry and electrochemical performance upon air ambient storage, thus helping to rationalize storage periods or air ingress during manufacturing.

We recently demonstrated that the APT analysis of pristine NMC811, following specimen preparation by focused-ion beam (FIB) milling at room temperature, was enabled by slight compositional changes of the surface when transferred through air, under ambient lab conditions[19], as showed in Figure 1a. The real space 3D atom map of the as-received pristine NMC811 in Figure 1b shows no substantial chemical fluctuation, as quantified by the composition profile in Figure 1c. Note that O atomic concentration is less than the nominal NMC composition since an oxygen molecule are formed due to the direct desorption and the neutral dissociation[21,22]. A series of similar APT specimens were exposed to air and imaged with the scanning electron microscope

(SEM) at intermediate times, Figure 1d and Figure S1. The change in contrast suggests a progressive change in the underlying microstructure. Upon exposure to ambient air for three months, protrusions on the side of the needle-shaped specimen formed, Figure 1e, precluding its analysis; it immediately fractured once we applied voltage on specimen due to the uneven surface morphology. APT analysis after 10-day exposure, Figure 1f–g, contrasts with the pristine material. The air exposure leads to a redistribution of Li and a change in the Li:Ni ratio towards the specimen's surface, either towards the apex, Figure 1g or towards the sides, Figure S2, with the formation of Li-depleted region, yet no carbonate species was detected, which indicates the layer is not a nickel carbonate species[4].

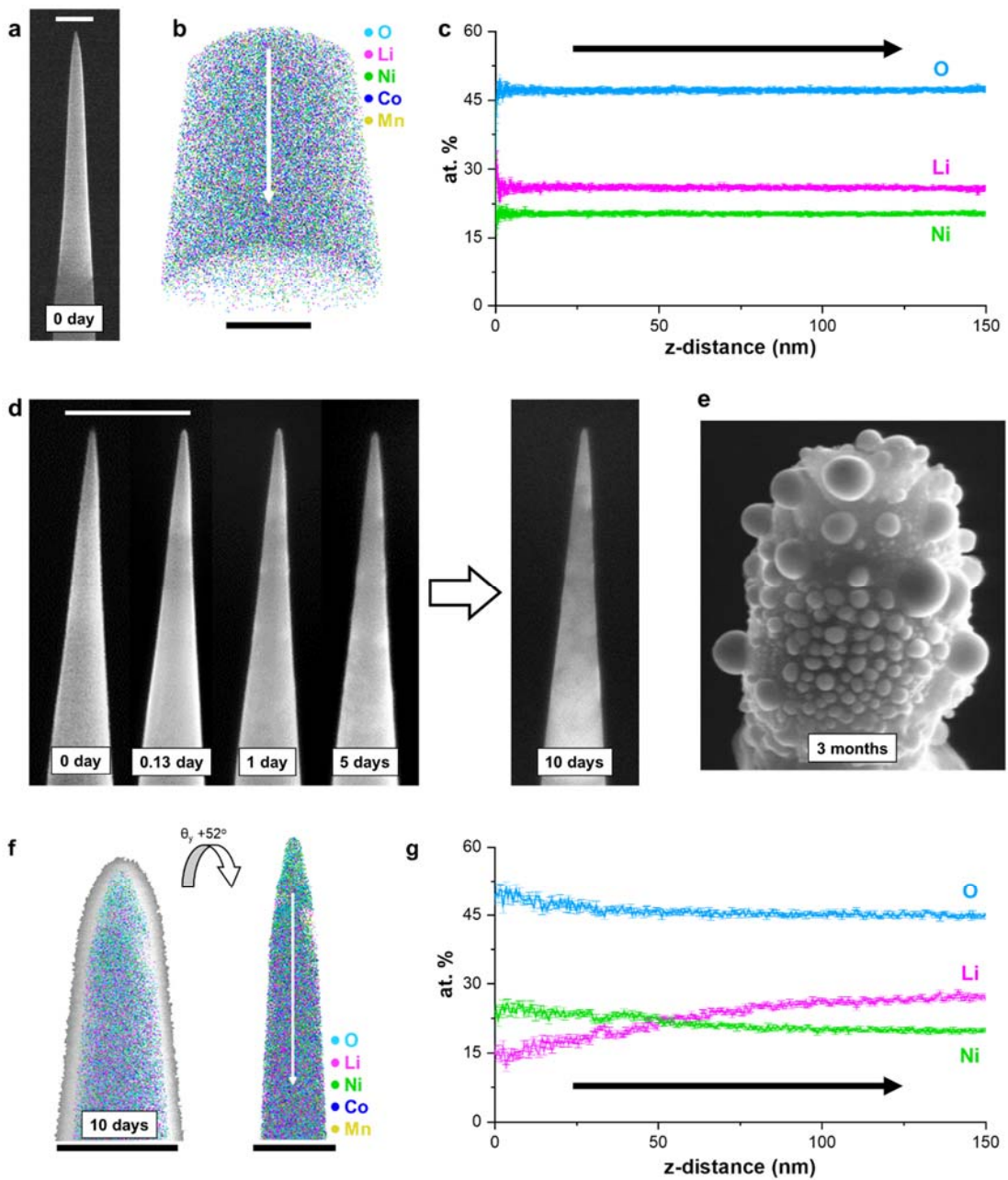

**Figure 1.** (a) SEM image and (b) 3D atom map of NMC811 APT specimen. (c) Corresponding 1D atomic compositional profile along the measurement direction (white arrow in Figure 1b). (d) Ex-situ oxidation experiment at different time span. (e) NMC specimen after three-month oxidation in ambient condition. (f) APT specimen after 10-days oxidation and electron-beam exposure for imaging. Inset image shows 3D atom map results with tilted at 52º in y-axis. (g) Corresponding 1D atomic compositional profile along the measurement direction (white arrow in Figure 1f). Co and Mn elements are not shown. Scale bars in SEM images are equal to 1 μm, whereas scale bars in atom maps are 50 nm.

This poses the important question of where does the surface Li go? Extreme caution is required when electron microscopy is used for characterizing Li materials as the electron beam can easily damage the structure and chemistry that could lead to mis-interpretation of results[13]. To assess the influence of the electron beam illumination on NMC811 during SEM imaging on the quantification of this elemental redistribution, we cross-sectioned an NMC811 particle with the FIB, and exposed it to air for a month. Even using a low acceleration electron-beam voltage of 5 kV, at relatively low electron-beam current of 0.6 nA, a short exposure of 5 sec in the SEM led to critical irradiation damage to the surface layer, Figures 2 and S3. The same behavior was observed when we illuminated the electron-beam on the surface-oxidized NMC 811 particle under TEM as it immediately reacted with the material (see Figure S4).

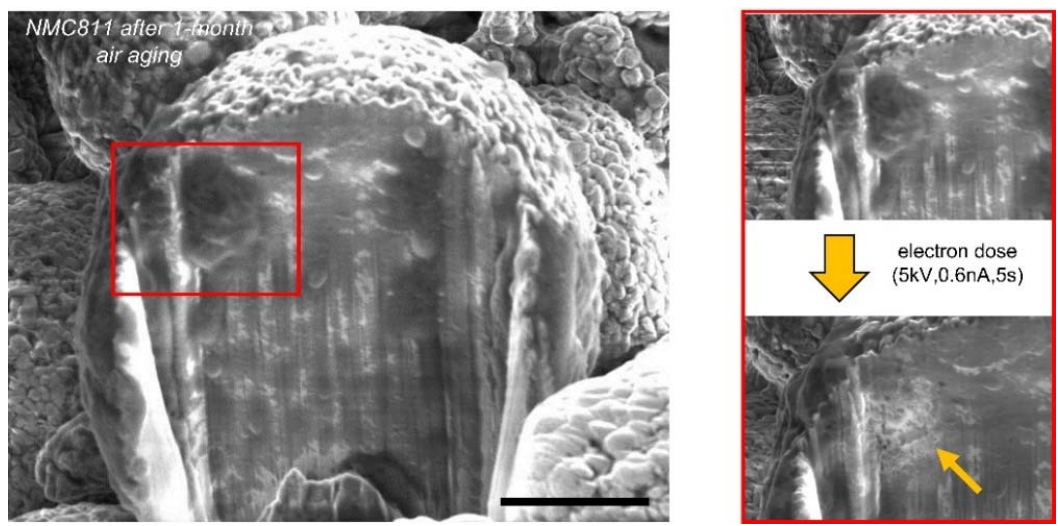

**Figure 2.** Electron beam-induced decomposition of $Li_2CO_3$ layers on air-aged NMC particle. The black scale bar is 5 μm. Yellow arrow remarks the deformation induced by electron-beams. Additional information is shown in Figure S3.

To analyze these beam-sensitive layers, we introduce a workflow that minimizes the degradation from the illumination by the electron beam necessary during imaging and specimen preparation.

We first cross-sectioned an NMC particle with the FIB, Figure 3a, during which electron beam usage was minimized and limited only to low magnification so as it is not concentrated in a certain region/particle. The cut NMC811 particles were then exposed to air for a month. A thick Cr-film (~200 nm) was then deposited by physical vapor deposition, as detailed in Figure S5-S7, in order to protect the reacted core-surface region from beam damage. Note that the Cr layer was thin enough to clearly discern the features of the particles, yet thick enough to protect their surfaces from electron-beam damage. APT specimens from these coated particles were then prepared at room temperature followed by sharpening at cryogenic temperature using the protocol outlined in Ref[23]. Figure 3b shows the corresponding 3D atom map, and underneath the Cr-cap is a carbon-rich region, with a ratio of Li to C of ~2, consistent with $Li_2CO_3$ (see composition in Table S1), followed by a gradient Li-depleted NMC region, as quantified by the composition profile in Figure 3c. As the $Li_2CO_3$ layer is very sensitive under electron-beam, it is easily removed when observed, hence it was not seen in the APT data after oxidation in Figure 1, and we were able to observe the remaining Li-depleted Ni-rich layer. The inhomogeneity of the surface structure of air-exposed NMC sample has not been previously reported and discussed in any APT analysis paper on Li-ion batteries[18,19,24–26]. This could be due to its elimination during imaging as previously mentioned, the limited field of view of current atom probe system[27], and the loss of the material during first laser alignment to the detector and the specimen[28]. Conversely, when the pre-run specimen was left in ambient air for 4 hr and resumed the measurement without electron-beam imaging, hydroxyl ions appear to be adsorbed on its surface[29].

Although substantial Li ions diffuse onto the NMC surface, the Li composition in the NMC does not vary much as the NMC is an abundant Li reservoir. Approx. 1 μm below the top surface, following further milling in the FIB after one-month oxidation, the elemental distribution is back to what was measured in the pristine sample (see Figure S9). The core of the sample shows a homogeneous elemental distribution with similar composition to the initial sample suggesting that the inner structure and chemistry won't be influenced much from air-oxidation in APT measurement.

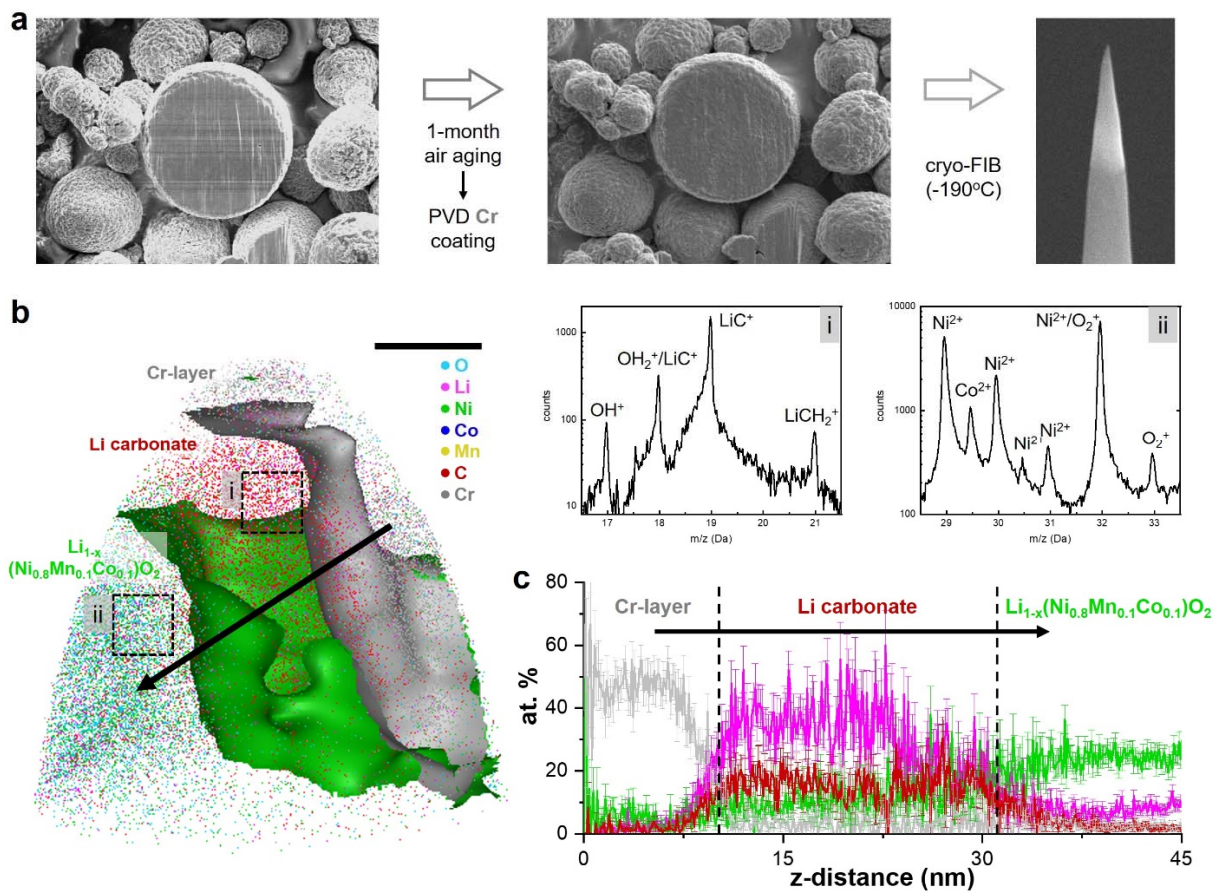

**Figure 3.** Atomic chemistry analysis of a month oxidized NMC particle. (a) The APT specimen preparation of surface-sensitive material with PVD coating and cryogenic milling. No electron beam imaging was done before the physical vapor deposition (PVD)-Cr coating and one-month ambient storage. (b) 3D atom map of Cr-coated oxidized NMC surface. Inset i and ii display extracted local mass spectra of interest regions in Figure 3b. In the extracted mass spectrum of the impurity layer, it indeed shows the peak of Li-C molecular ions that we can conclude the presence of $Li_2CO_3$. The black scale bar is equal to 10 nm. (c) 1D compositional profile of elements across the oxidized NMC surface (cylindrical region of interest of $\Phi 5 \times 50$ nm$^3$). Co, Mn, and O elements are not plotted. The details of the mass spectrum analysis are presented in Figure S8.

In the reactive environment, the strong affinity between Li and gases in the reactive environment results in Li-ion migration towards the surface where it reacts with gaseous species. In-situ, environmentally-controlled TEM recently revealed a successive reaction of NMC cathode materials with water, leading to the formation of a LiOH layer on the surface[30]. A strong driving force for Li migration to form LiOH explains the Li concentration gradients found in Figure 3c. LiOH spontaneously reacts with $CO_2$ at room temperature to form $Li_2CO_3$ ($\Delta G$ = -88 kJ/mol)[31]. This reaction explains the use of LiOH canisters for $CO_2$ removal in space shuttles[32]. However, $Li_2CO_3$ is a wide bandgap insulator that acts as a passivation layer to slow the electrode-electrolyte because of sluggish electrons transport. It exhibits extrinsic conduction, essentially by electro-mobility of interstitial Li ions[33], yet it can act as a barrier against chemical delithiation, and its presence results in inferior reversibility and cyclability[34].

Pre-existing intra-granular cracks from the synthesis have been considered to be one of the major degradation mechanisms[35,36], since they generate build-ups of large strains during the phase transition and the chemical reaction with electrolyte[9,35,37]. We exposed a crack-containing NMC811 particle to air and, in Figure 4a, the red arrow indicates a region where the crack swelled, while a different phase has grown on its surface that the APT analysis reveals to be also $Li_2CO_3$. When exposed to a reactive environment, the fresh surface created by the propagating crack triggers further phase transformation into a carbonate, which induces Li diffusion and consequently leads to locally lower electrical conductivity and internal Li-ion mobility, accelerating degradation of the battery life and its performance[35].

Our findings point to four distinguishable layers formed on the Ni-rich oxide cathode particle stored in ambient conditions: a Li-carbonate, a Li-depleted layer, likely to be a 2-nm-thick rock-salt shell previously observed by in-situ TEM[30], a lithium diffusion (Ni-rich) layer, and finally the pristine NMC811 phase. Each region will have a different composition and possibly structure, and hence impedance, which, therefore, must be considered to fully comprehend the critical capacity loss of Ni-rich cathode materials. Figure 4c provides a schematic of inhomogeneous Li distribution from the formation of different phases under air exposure and their respective resistivity.

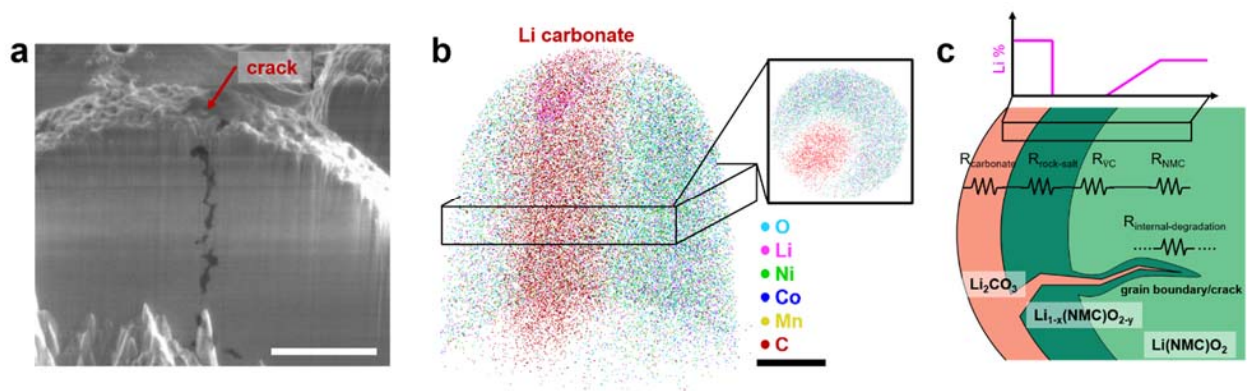

**Figure 4.** (a) Cross-sectioned air-stored NMC particle (scale bar = 5 μm). (b) APT result of the NMC with $Li_2CO_3$ (scale bar = 20 nm). (c) A schematic illustration of NMC 811 material after air-oxidation with a Li concentration profile along the surface. Additionally, Figure S10 shows many air-susceptible sites for NMC particles: grain boundaries, cracks, triple junction, surface etc.

In summary, we systematically investigated the atomic scale compositional evolution of Ni-rich NMC811 cathode powder exposed to ambient air. By adopting a cryogenic APT specimen preparation technique for the electron-beam sensitive Li compounds, we carefully confirm Li-ion extraction from the particles driven by the formation of LiOH that reacts with $CO_2$ in air to form $Li_2CO_3$ and creates a local Li and Ni concentration gradient along the surface. Beyond the surface of individual particles, the surface of cracks from synthesis or arising from the volume change during the charge/discharge cycles, are also covered with $Li_2CO_3$, thereby increasing internal stresses and further contributing to facilitate the nucleation and/or growth of stress corrosion cracks.

Although the changes we image are on a scale of only tens of nanometers, the associated differences in Li-ion mobility and impendence can induce high initial voltage peak upon delithiation in the first charge and irreversible capacity loss[4,38]. Alleviating the formation of different phases should be prioritized, thus minimizing the inhomogeneity in the Li distribution before and during cycling, so as to retain a stable charge and discharge performance assessment.

**Acknowledgements**

SHK and BG are grateful for financial support from the ERC-CoG-SHINE-771602 at some point in the past few years. SHK and BG acknowledge financial support from the DFG through the DIP Project number 450800666. XZ is grateful for financial support from the Alexander von Humboldt Foundation.

# Atomic scale evolution of the surface chemistry in NiMnCo cathode for Li-batteries stored in air

Mahander P. Singh[a], Se-Ho Kim[a,*], Xuyang Zhou[a], Hiram Kwak[b], Stoichko Antonov[c], Leonardo Shoji Aota[a], Chanwon Jung[a], Yoon Seok Jung[b], Baptiste Gault[a,d,*]

a Max-Planck-Institut für Eisenforschung, Düsseldorf, Germany

b Department of Chemical and Biomolecular Engineering, Yonsei University, Seoul, South Korea

c National Energy Technology Laboratory, Albany, Oregon, USA

d Department of Materials, Royal School of Mines, Imperial College London, London, UK

* co-corresponding authors: s.kim@mpie.de; b.gault@mpie.de

**Methods**

*Atom-probe specimen preparation and air oxidation*

Ga-ion focused-ion-beam (FEI Helios 600 (Thermo-Fisher)) was used to fabricate needle-like APT specimens from the pristine NMC811 particles following the standard lift-out method. The final milling protocol was performed at the ion-beam accelerated voltage of 5kV and current of 8pA to remove the residual Ga contaminants. For storage, as-prepared APT specimens were exposed in typical ambient conditions in Düsseldorf (average relative humidity of 77% and $CO_2$ level of 413 ppm). The SEM images were taken immediately after air oxidation of the APT NMC811 specimens at certain time intervals. After a 10/30 days oxidation, the specimen was transferred through air and loaded into the atom probe analysis chamber.

*Horizontal-cut NMC811 and PVD-coating process*

The pristine NMC811 particles were mounted on a 52º tilted stage (see inset in Figure S6) then the stage was tilted at 52º so that the particles of interest were aligned to the ion beam column for a horizontal cut. The selected particles were sectioned using Ga ion beam voltage of 30kV and current of 9nA. To eliminate influence of residual Ga during the surface oxidation, the ion-beam cleaning process was carefully performed at condition of 5kV&15pA, 5kV&8pA and 2kV&10pA each for 60s. After one-month storage at ambient conditions, a Cr thin film was deposited to protect the surface of the oxidized NMC811 powder. The thin film was grown to a thickness of approximately 200 nm at room temperature at a rate of 0.12 nm/s by magnetron sputtering of high purity Cr (99.95%) in a direct current (DC) gun in a physical vapor deposition (PVD) cluster (BESTEC, Berlin, Germany). Prior to sputtering, the chamber was pumped to a base pressure of $1 \times 10^{-5}$ Pa. Argon was then introduced as the working gas at a flow rate of 40 standard cubic centimeters per minute (SCCM) to achieve a pressure of 0.5 Pa (see Figure S5).

*Cryo-APT specimen preparation*

The Cr-coated oxidized particles were prepared into APT specimen. First, clean tranches were milled on the coated surface for the lift-out process. To prevent the contamination of platinum deposition precursor, no Pt/C passivation layer was deposited on top of the sample's surface. A wedge-shaped lamella was mounted on a pre-sharpened Si microtip. To avoid e-beam damage, a rapid scanning flash image was taken for each of the successive steps of the pillar preparation (see Figure S7a-S7d). Subsequently, the cryo-stage was implemented to avoid uncontrollable ion/electron beam damage of the sensitive oxidized layer during the annular milling process. The cryo-stage (Gatan C1001, Gatan Inc) was pre-cooled to -190ºC by cold gaseous $N_2$ externally connected to the FIB chamber. The ion beam voltage of 30kV and current of 80pA were used for annular milling and progressively smaller milling patterns were set to sharpen the Cr/NMC811 interface (see Figure S7e-S7h). After the final milling process, the cryo-as-prepared specimens were transferred from the FIB chamber to the LEAP atom probe system.

*APT Characterization*

APT measurements were performed with Cameca LEAP 5000 XS system in pulsed UV laser mode at a detection rate of 0.5%, a laser pulse energy of 1-10 pJ, and a pulse frequency of 100 kHz. The specimen temperature was set to 60 K throughout the analysis. Data reconstruction and analyses were performed using the commercial software AP SUITE 6.1 developed by Cameca Instruments.

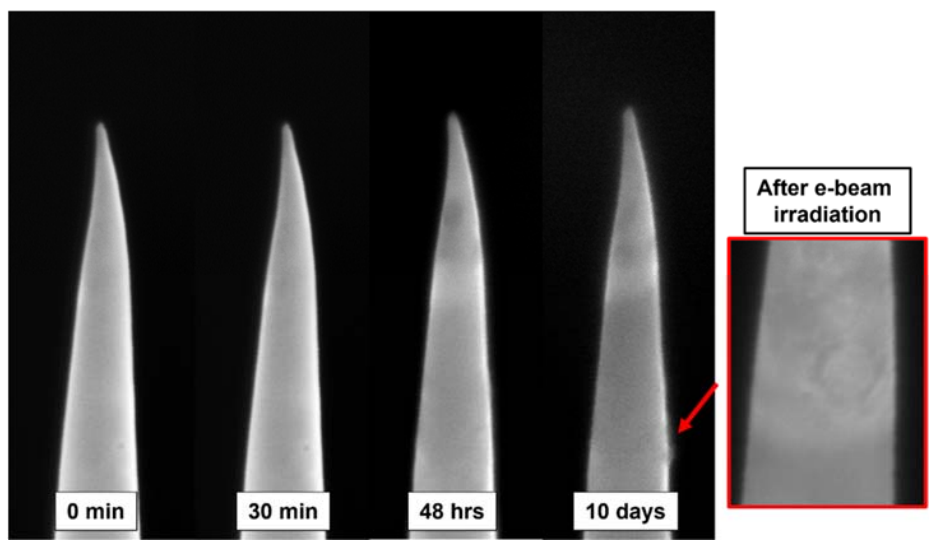

**Figure S1.** Microstructure evolution over air exposure. Note that e-beam was irradiated on $Li_2CO_3$ layers as observed by backscattered electrons.

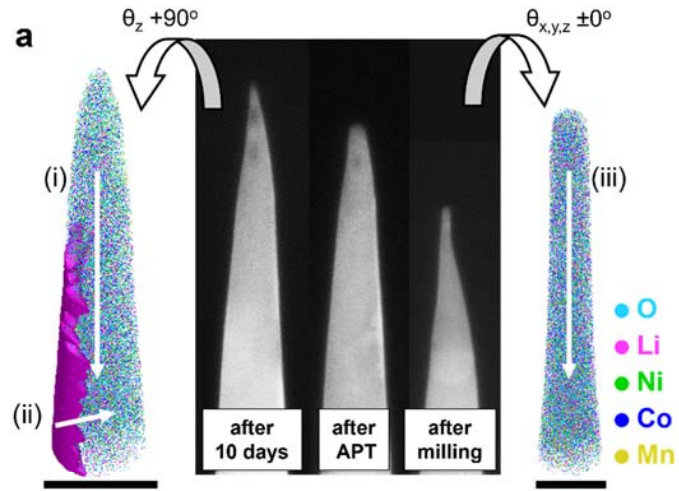

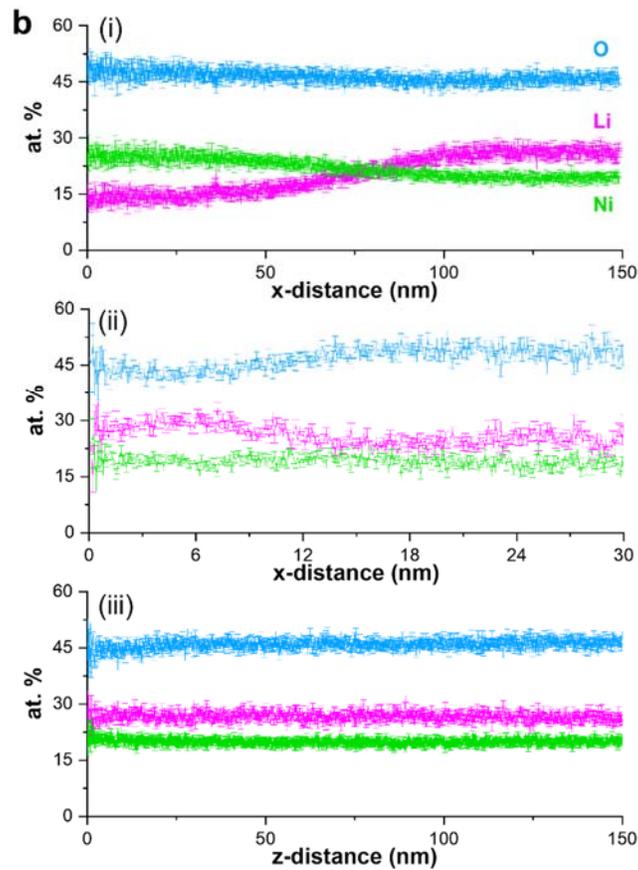

**Figure S2.** (a) APT specimen after 10-days oxidation and e-beam exposure for imaging. Inset image shows 3D atom map results with tilting at 52° in y-axis. Iso-surface of Li at 25 at.% displays the interface

of different Li content phases. All scale bars are 50 nm. (b) Corresponding 1D atomic compositional profile along the measurement direction (white arrow in Figure S2a). Co and Mn elements are not shown.

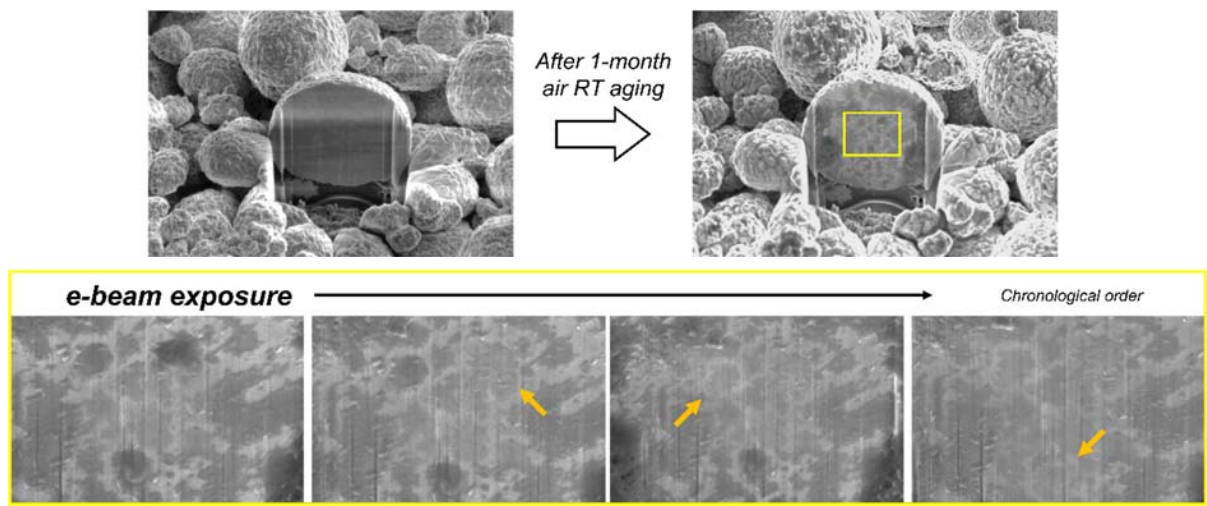

**Figure S3.** Electron beam-induced decomposition of $Li_2CO_3$ layers on the air-aged NMC particle under SEM.

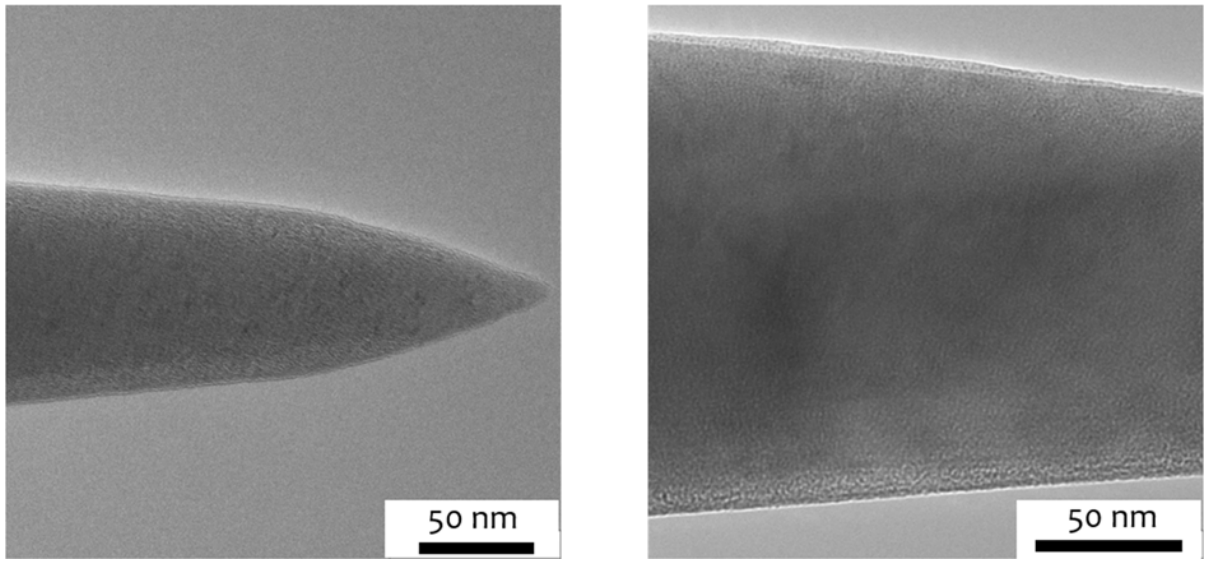

**Figure S4.** Beam-induced decomposition of $Li_2CO_3$ layers on the air-aged NMC particle under TEM.

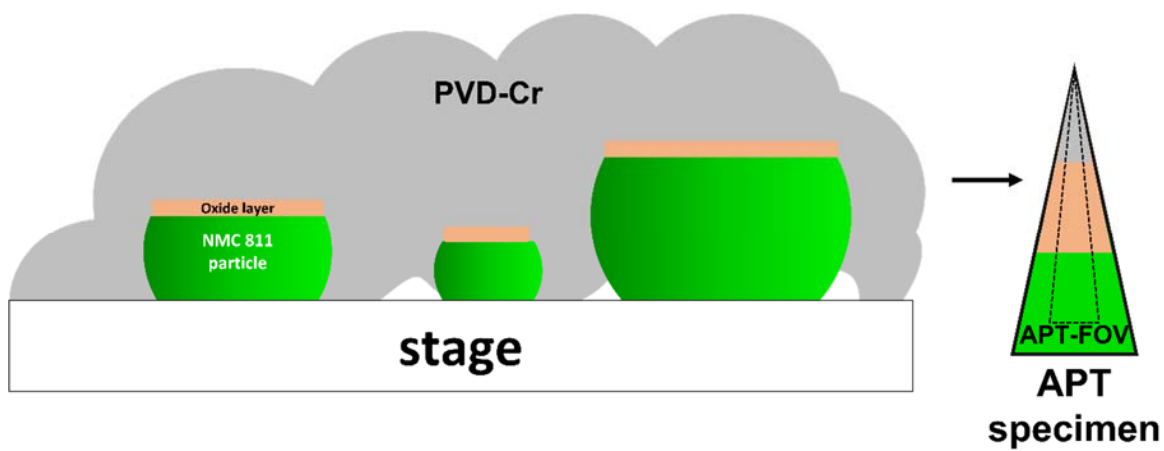

**Figure S5.** A schematic illustration of oxidized layer APT sample preparation. No SEM image was taken between oxidation and coating step.

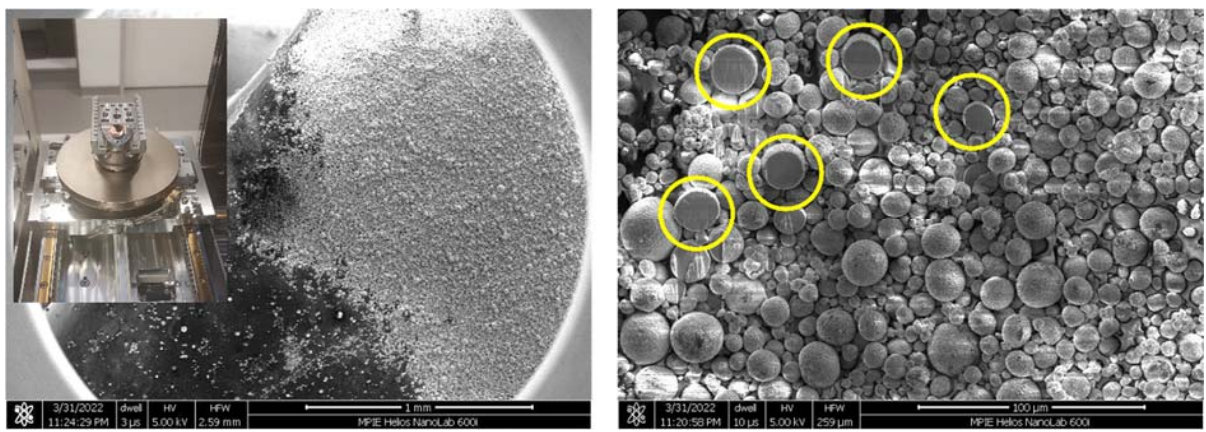

**Figure S6.** (a) NMC811 particles mounted on 52º-tilted stage holder. (b) horizontally sectioned particles for cryo-APT experiment.

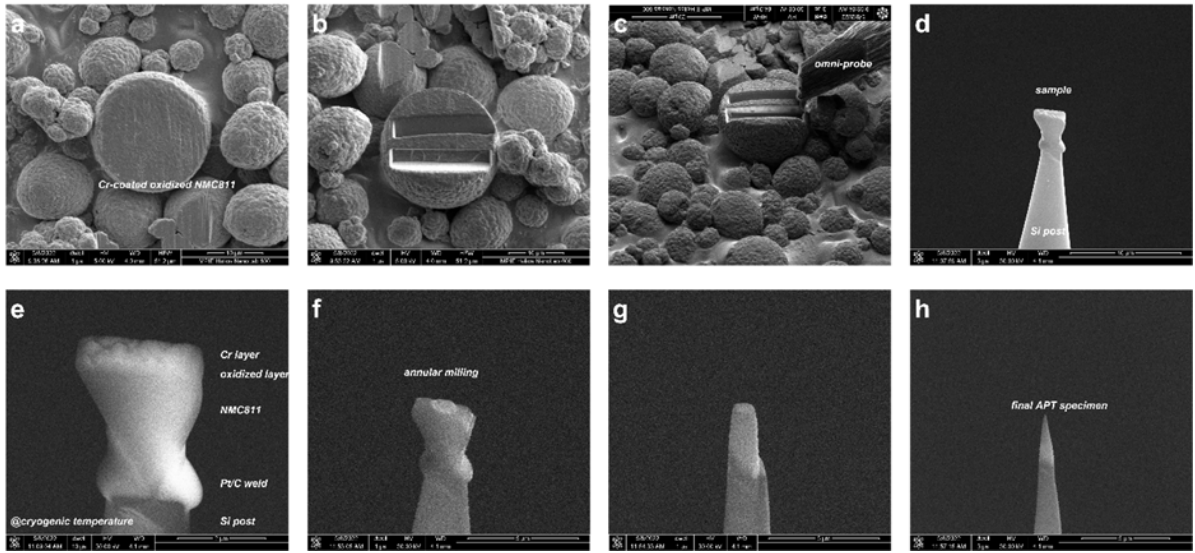

**Figure S7.** (a) Horizontally sectioned NMC811 particle coated with Cr film after a month ambient storage. (b) Wedge-shape cut for lift-out process. (c) A micro-manipulator was inserted to free the lamella. (d) The pillar of NMC811 on a Si microtip. (e) Cooled APT sample before annular milling. (f)&(g) Annular milling process. (h) Final APT specimen.

**Table S1.** Atomic composition of extracted cubic region of interest (i) in Figure 3b. Note that O content is low because of the neutral oxygen gas formation when it desorbed from the specimen apex.

| Elements | Li | O | C | H | Ni | Cr | Co | Mn | Ga |
|---|---|---|---|---|---|---|---|---|---|
| **at.%** | 38.670% | 21.791% | 19.466% | 11.345% | 6.183% | 1.950% | 0.335% | 0.199% | 0.060% |

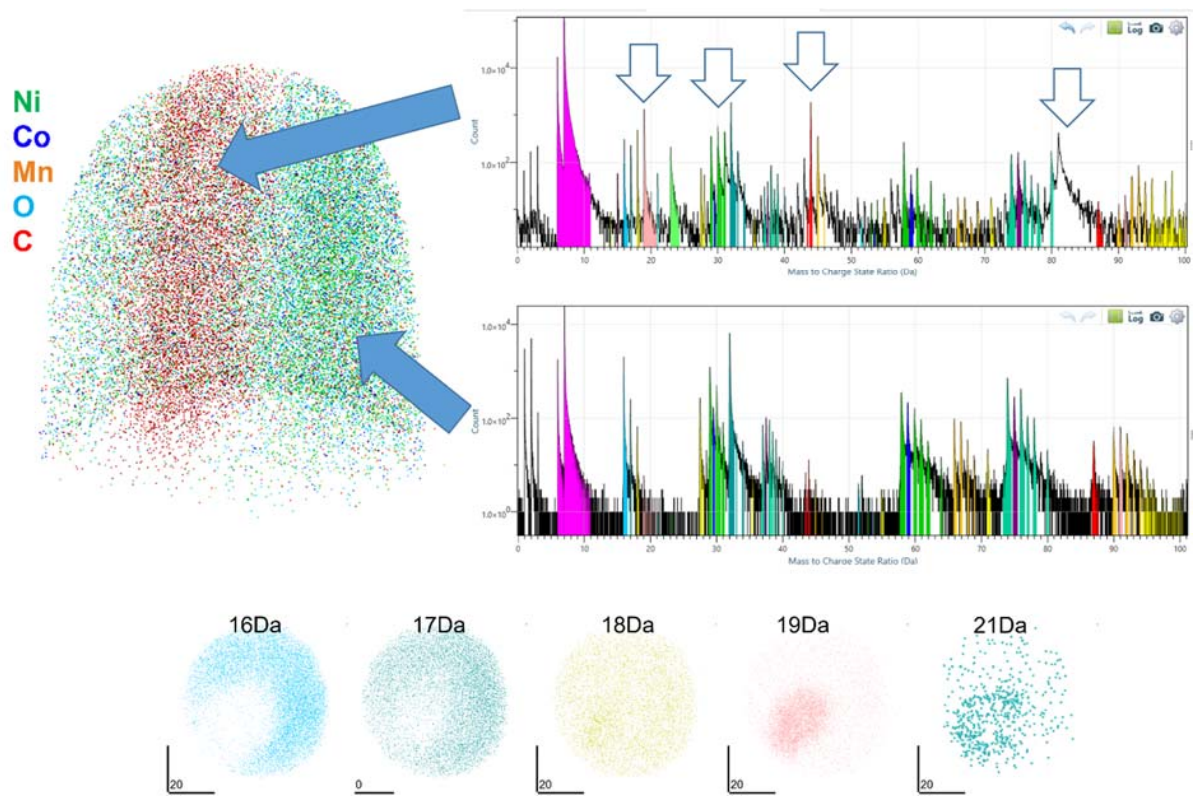

**Figure S8.** APT result of oxidized crack region. Mass spectra analyses indicate that the C and O originating peaks were measured at crack/oxidized region. The peaks at 16, 17, 18, 19, and 20 Da correspond to $O^+$, $OH^+$, $OH_2^+$, $LiC^+$, and $LiO^+$, respectively.

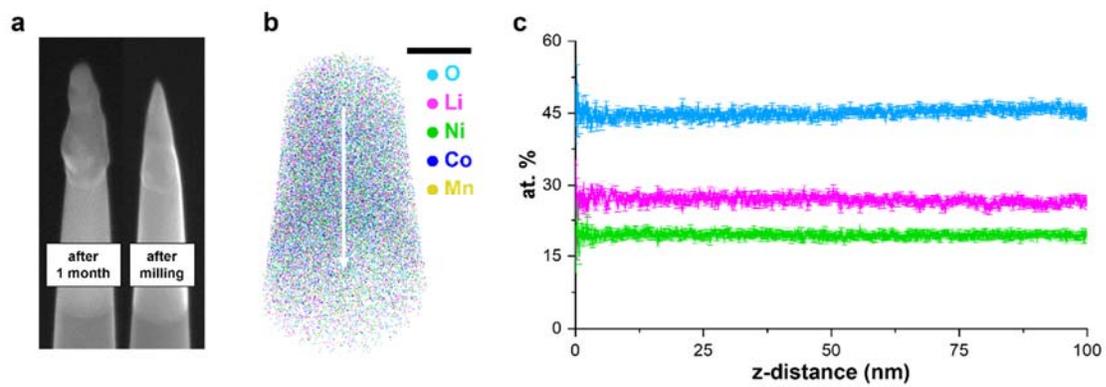

**Figure S9.** (a) APT specimen after a month oxidation and after ion milling. (b) APT measurement of the one-month-oxidized re-sharpened specimen.

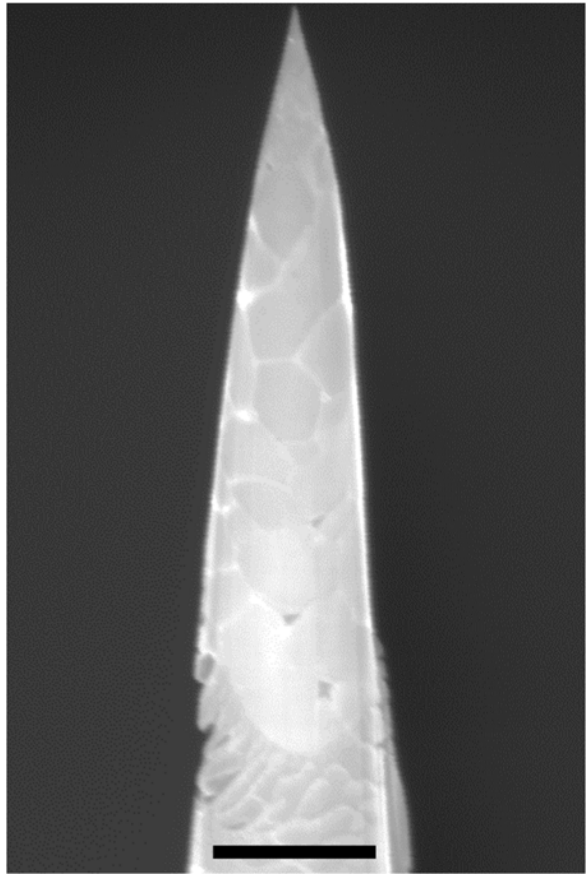

**Figure S10.** An example of NMC811 APT specimen from the cryo-preparation. A scale bar is 1μm.